# Magnetic torque anomaly in the quantum limit of the Weyl semi-metal NbAs

Philip J.W. Moll[1], Andrew C. Potter[1], Brad Ramshaw[2], Kimberly Modic[2], Scott Riggs[3], Bin Zeng[3], Nirmal J. Ghimire[4], Eric D. Bauer[4], Robert Kealhofer[1], Nityan Nair[1], Filip Ronning[4], and James G. Analytis[1]

[1]Department of Physics, University of California Berkeley, CA, USA
[2]National High Magnetic Field Laboratory, Los Alamos, NM, USA
[3]National High Magnetic Field Laboratory, Tallahassee, FL, USA
[4]Los Alamos National Laboratory, Los Alamos, NM, USA

*correspondence address: moll@berkeley.edu

**Electrons in materials with linear dispersion behave as massless Weyl- or Dirac-quasiparticles, and continue to intrigue physicists due to their close resemblance to elusive ultra-relativistic particles as well as their potential for future electronics[1–3]. Yet the experimental signatures of Weyl-fermions are often subtle and indirect, in particular if they coexist with conventional, massive quasiparticles. Here we report a large anomaly in the magnetic torque of the Weyl semi-metal NbAs upon entering the "quantum limit" state in high magnetic fields, where topological corrections to the energy spectrum become dominant. The quantum limit torque displays a striking change in sign, signaling a reversal of the magnetic anisotropy that can be directly attributed to the topological properties of the Weyl semi-metal[4–6]. Our results establish that anomalous quantum limit torque measurements provide a simple experimental method to identify Weyl- and Dirac- semi-metals.**

The observation of electrons behaving as Dirac- and Weyl-fermions in condensed matter[1] systems such as graphene[7], surface states of topological insulators[8], Dirac semi-metals such as $Na_3Bi$[9,10], $Cd_3As_2$[11,12] and the mono-pnictides class of Weyl semi-metals (Ta,Nb)(As,P)[13–15] has sparked considerable excitement. The research interest is partially fueled by the technological potential of these materials in devices exploiting the relativistic nature of the electrons, such as ultra-high mobilities, the topological protection from back-scattering and the massless behavior of charge carriers. In addition, these material classes give us the opportunity to test concepts and predictions that have been put forward to describe massless fermions in high energy physics, for example the "chiral anomaly"[14,16]. Given the wealth of novel phenomena in these materials, it is important to find clear experimental signatures that can help identify new compounds as Weyl and Dirac systems. Here we show that the non-zero Berry's phase in Weyl semi-metals manifests itself as a magnetic torque anomaly at the quantum limit. Strong magnetic fields quantize the electronic motion perpendicular to the field onto Landau levels of energy;

$$\varepsilon_{n,k} = \begin{cases} \frac{\hbar eB}{m_{eff}}(n+\gamma) + \frac{\hbar^2 k_z^2}{2m} & \text{Trivial metal} \\ \hbar v \sqrt{2B(n+\gamma) + k_z^2} & \text{Weyl metal} \end{cases} \quad (1)$$

, where $e$ denotes the electron charge, $n$ the Landau level index, $m_{eff}$ the effective electron mass, $v$ the Fermi velocity, and $k_z$ the momentum component along the magnetic field B. In usual metals, the quantum correction term $\gamma$ takes the value ½[17], but in Weyl systems it attains a contribution known as Berry's phase[5,18], such that $\gamma = 0$. This is a topological property that depends only on the existence of Weyl nodes and not on the details of the band structure.

One obvious consequence is the behavior in the last Landau level (n=0): While these states in a normal metal continue to gain energy linearly with increasing field strength, those in Weyl-metals become independent of the magnetic field ($n, \gamma = 0$). The magnetization per electron of a normal metal at zero temperature in the quantum limit saturates to a field independent value $M_{n=0} = -\partial E_0/\partial H = -\frac{\hbar e}{m_{eff}}\gamma$, while in the Weyl case the magnetization of the conduction electrons vanishes, $M_{n=0} = 0$. While the saturating magnetization of normal metals yields a smooth evolution of the torque at the quantum limit, the collapse of the magnetization in Weyl metals leads to a torque anomaly[19,20]. Similar argument have explained an anomalous magnetization of LaRhIn$_5$ in the quantum limit, where a different type of topological defect (band-crossing line) generates Berry's flux[21–23].

NbAs was chosen as a candidate Weyl semi-metal to search for the magnetic anomaly at the quantum limit[24]. The small Fermi surfaces of this material allow us to reach and exceed the quantum limit for a large angle range in static magnetic fields up to 45T and pulsed magnetic fields up to 65T. Similar to the recently investigated related phosphides NbP and TaP, NbAs is expected to host both trivial electrons and Weyl fermions on different Fermi surface sheets[13,25]. We have measured resistance and magnetic torque of single crystals of NbAs using both capacitive and piezo-resistive cantilever torque techniques in pulsed and dc-fields, and find quantitative agreement between different samples (see Methods). The magnetic torque $\tau = M(H) \times H$ is a direct measure of the magnetic anisotropy of a crystal $(\chi_\perp - \chi_\parallel)$ and as such most sensitive to even small changes in the magnetic response at high fields.

Figure 1 contrasts the torque and resistivity of NbAs in high fields, highlighting the main experimental observation: at the quantum limit, the magnetic torque shows a well-defined kink and a cross-over to a sizeable, linear increase without a sign of saturation up to the highest accessible fields in this experiment of 65T. The position of the break in slope in the torque around 20T (for fields 25° off the c-axis towards the a-axis) matches well the observed extremal Fermi surface cross-section from the low-field quantum oscillation measurements and the position of the last quantum oscillation of the resistivity, providing independent confirmation that the torque anomaly coincides with the quantum limit. Two close-by frequencies, F1 and F2, are observed in NbAs, in agreement with band structure calculations predicting the presence of both trivial and non-trivial Fermi surface sheets[14]. The limited field window prevents a clear identification of the Fermi surface responsible for the torque anomaly, yet a previous quantum oscillation study associated the smaller frequency, F1, with Weyl fermions[26].

The transverse magnetoresistance is very large, similar to recent reports in topological metals[14,15,27], reaching a $\frac{\Delta R}{R} \sim 9320$ from $\rho(0T) = 0.3 \mu\Omega cm$ to $\rho(45T) = 2833.3 \mu\Omega cm$. Yet the smoothly saturating resistivity remains featureless as the system crosses the quantum limit, in strong contrast to the torque.

The Fermi surface anisotropy leads to an angle dependence of the quantum limit field (Fig 2). The extremal cross-sections of the Fermi surfaces grow from 16T for fields aligned with the crystal c-axis to 80T as the field is tilted towards the a-axis. The position of the kink in the magnetic torque is found to coincide with the quantum limit field obtained by dHvA frequencies over the whole angle range up to 65° (Fig 2c), above which the quantum limit exceeds the maximal fields available in this experiment of 65T. The break in slope of the magnetic torque thus is clearly associated with the transition into the ultra-quantum limit (n=0).

In addition to the distinct change in behavior above and below the quantum limit, the torque also changes sign. A reversal of the sign of the magnetic torque $\tau \propto (\chi_\perp - \chi_\parallel) \sin(2\theta)$ indicates that either the anisotropy or the sign of the induced magnetization is inverted in high-field state of NbAs compared to its low field value. Figure 3 shows the angle dependence of the torque at fixed fields, in the high-field (60T) and low-field (10T) region as well as an intermediate field value (30T). Both the high- and low-field torque follows a $\sin(2\theta)$ dependence, as expected for metals without permanent magnetic moments. The phase of these sines, however, is changed by $\pi$, which again shows the reversal of the anisotropy between the high- and low-field state. This reversal occurs directly at the quantum limit, as evidenced by the strong deviation from $\sin(2\theta)$ of the torque at intermediate fields. At low angles close to the c-axis and fields of 30T, the system is above the quantum limit (compare with Fig 2). As the field is rotated towards the a-axis, the torque increases tracking the high-field behavior. At the same time the quantum limit field increases upon tilting the field towards the a-axis due to the anisotropy of the Fermi surface, and reaches 30T at around 45°. At this point, the system crosses into the low-field state as the quantum limit is pushed above 30T, which is accompanied by a sudden drop of the torque crossing through zero and reaching the same amplitude on the negative side. As the angle is increased further, the torque now tracks the low-field behavior. This firmly establishes that an abrupt reversal of the magnetic anisotropy occurs at the quantum limit.

To model this torque reversal, we begin by computing the magnetization for an idealized isotropic Weyl node: $M_{iso}(H) = -\frac{dE_0}{dH}$, where $E_0(H) = \frac{eH}{2\pi\hbar c}\sum_{n,k}\text{sgn}(n)\varepsilon_{n,k} n_F(\varepsilon_{n,k} - \mu)$, is the ground state energy, and the Fermi-function $n_F$ restricts to occupied orbitals. The chemical potential, $\mu$, is determined by the total particle density, $\rho$, (measured relative filling up to the Weyl node) via the relation: $\rho = \frac{H}{2\pi}\sum_{n>0,k} n_F(\varepsilon_{n,k} - \mu)$. The valence band states (n<0) are far from the quantum limit and contribute a non-oscillatory diamagnetic response $M_{n<0} \approx -\frac{vH}{4\pi^4\hbar c}\log\frac{\Lambda^2}{H}$, where $\Lambda$ is the characteristic momentum at which the dispersion deviates from its linear relativistic form. The low-density conduction states ($n > 0$) initially contribute a paramagnetic response that dies away for increasing field as the occupied conduction states are subsumed into the $n = 0$ level, whose energy is field-independent. To model the torque measurements, the results of this simplified isotropic model can be related to the more realistic case with anisotropic velocities along the c- and a,b- axes: $v_c = \lambda v_{a,b}$ by a simple rescaling:

$$\boldsymbol{M}(H,\theta) = \frac{\lambda^{-2}\cos\theta\,\hat{x} + \sin\theta\,\hat{z}}{\sqrt{\cos^2\theta + \lambda^2\sin^2\theta}} M_{iso}\left(H\sqrt{\cos^2\theta + \lambda^2\sin^2\theta}\right) \quad (3)$$

We note that comparing the quantum-limit field for $\theta = 0°, 90°$ from the quantum oscillation spectrum (Fig. 2c) yields an anisotropy parameter $\lambda \approx 0.4$. The sign reversal of magnetization and torque due to the conduction bands entering the quantum limit matches that observed experimentally. The residual diamagnetic contribution beyond the quantum limit comes from the valence band electrons, and will be enhanced by additional pockets of massive non-relativistic electrons – if present as in the case of NbAs (Figure 4).

This torque reversal at the quantum limit is a direct consequence of the Berry's flux contained in the Fermi surface and does not depend on the details of the band-structure, and as such is linked to the non-trivial topology of Weyl semi-metals. As the quantum limit is in experimental reach for the small Fermi surfaces around Weyl- and Dirac-points, high field torque measurements provide a promising experimental tool to identify new compounds as topological metals.

## Methods

**Crystal Synthesis** - NbAs single crystals were grown by chemical vapor transport using iodine as the transport agent. First, NbAs powder was prepared by heating a stoichiometric mixture of Nb powder and As pieces sealed in a quartz tube under vacuum. The ampule was slowly heated up to 700°C and kept at this temperature for 3 days. Then 2 grams of NbAs powder was mixed with 0.5 gram of iodine and sealed in a quartz tube. The sealed ampule was loaded into a horizontal tube furnace for 10 days. The temperature of the hot end was maintained at 950°C and that of the cold zone was approximately 850°C. Several well facetted crystals were obtained inside the quartz tube. The crystal structure was verified using room temperature x-ray diffraction.

**Torque Measurements** - The magnetic torque was measured at the National High Magnetic Field Laboratory using CuBe cantilevers with capacitive readout in steady fields up to 45T, and in pulsed fields up to 65T using piezoresistive cantilever (SEIKO-PRC120). As the torque becomes substantial beyond the quantum limit, the resulting large deflection angle can drive both techniques out of their linear response regime. To correct for deviations from non-linearity, field-sweeps of positive and negative polarity were averaged.

## Acknowledgements

We thank Itamar Kimchi for helpful discussions. A.C.P. was supported by the Gordon and Betty Moore Foundation's EPiQS Initiative through Grant GBMF4307. Torque measurements were supported by the Gordon and Betty Moore Foundation's EPiQS Initiative through Grant GBMF4374. N.G., E.D.B, and F.R. were supported under the auspices of the Department of Energy, Office of Basic Energy Sciences, Division of Materials Science and Engineering. Work at NHMFL-LANL is carried out under the auspices of the National Science Foundation, Department of Energy and State of Florida.

## Author contributions

P.J.W.M, F.R. and J.G.A designed the experiment. P.W.J.M, B.R., K.M, S.R.,B.Z & N.N. performed the high-field measurements, N.N. the low-field SdH experiments. N.J.G. and E.D.B. grew and characterized the NbAs crystals. A.C.P. developed the theoretical model. P.J.W.M., A.C.P., F.R. and J.G.A. wrote the manuscript. B.Z. acknowledge support from the

# Figures

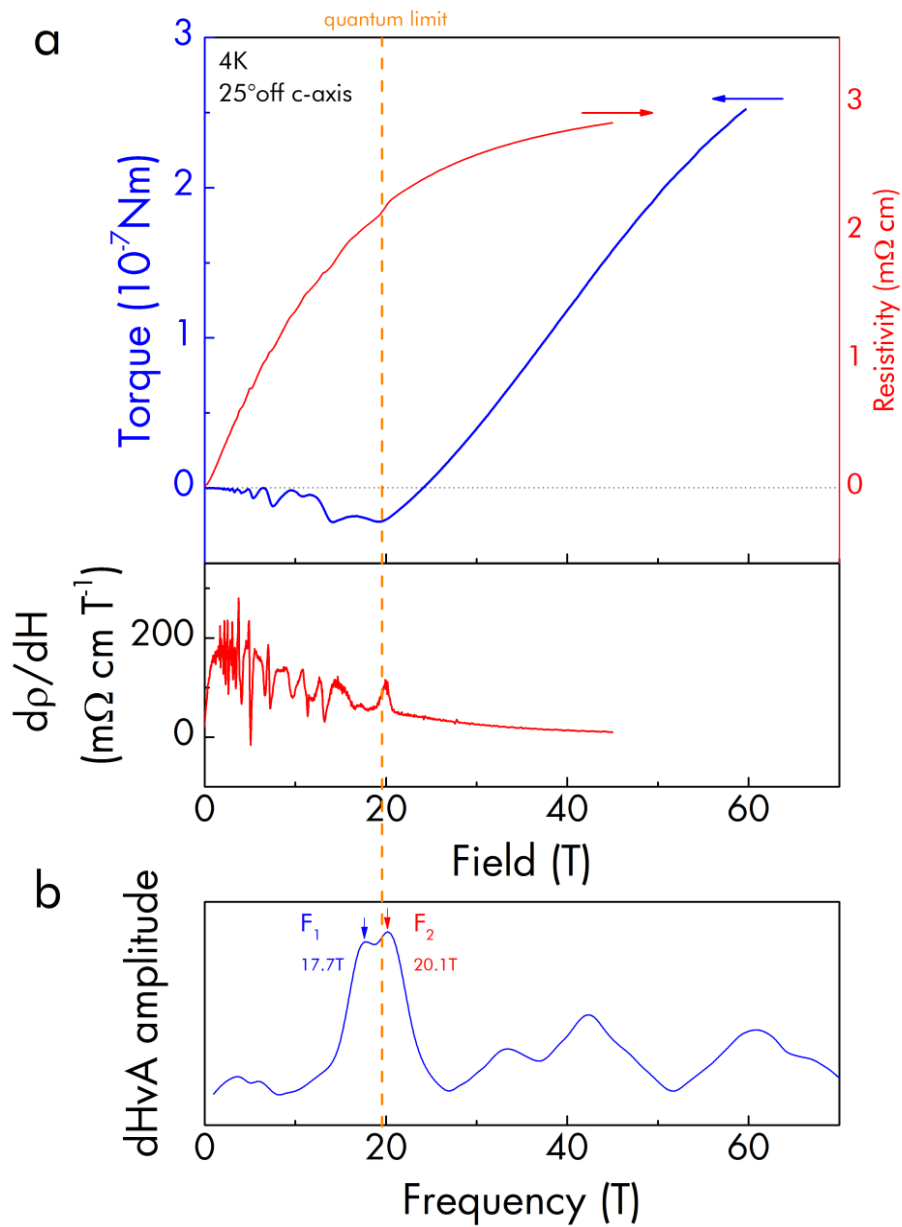

**Fig. 1: Magnetic Torque and resistivity of NbAs across the quantum limit.** (a) The low-field torque follows the conventional quadratic behavior and shows strong quantum oscillations up to the quantum limit (orange line). At higher fields, the torque grows strongly in magnitude and is approximately linear in field and crosses zero, signaling a reversal of the magnetic susceptibility. In contrast to the torque, there is no change in behavior in the smoothly saturating magnetoresistance beyond the last quantum oscillation. The quantum limit field is self-consistently determined through the analysis of the de Haas-van Alphen oscillation, indicating two close-by extremal orbits F1 and F2.

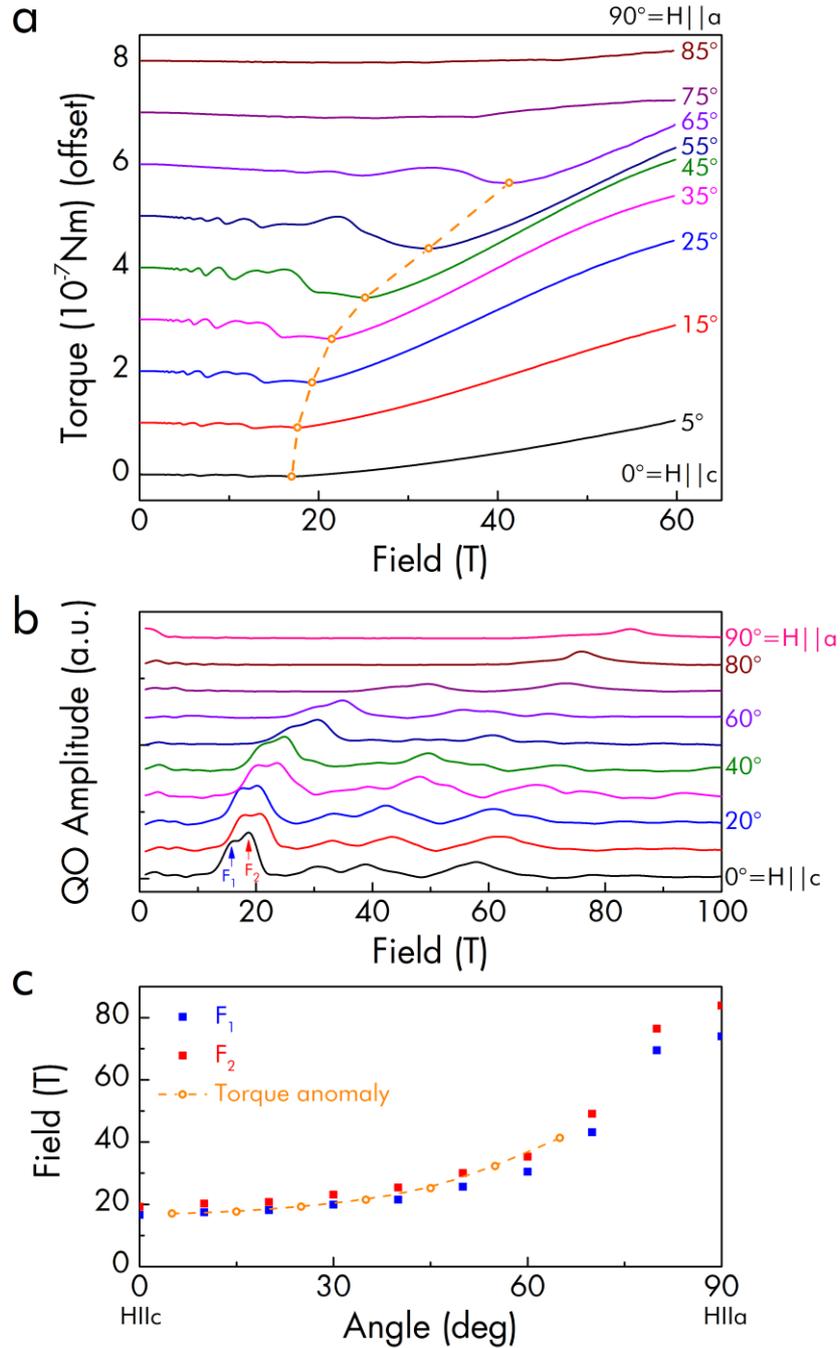

**Fig. 2: NbAs magnetic anomaly.** (a) Angle dependence of the torque in high magnetic fields at 4K, tilting the field from the c-direction (0°) towards the a-direction (90°). A clear break in slope at the quantum limit is observed in all traces up to 65°. (b) The quantum limit as a function of field angle was determined by Shubnikov-de Haas oscillations. The oscillations contain a large harmonic contribution, leading to pronounced higher harmonics in the FFT spectrum. (c) The quantum limit tracks well the torque anomaly at all angles where it is experimentally accessible.

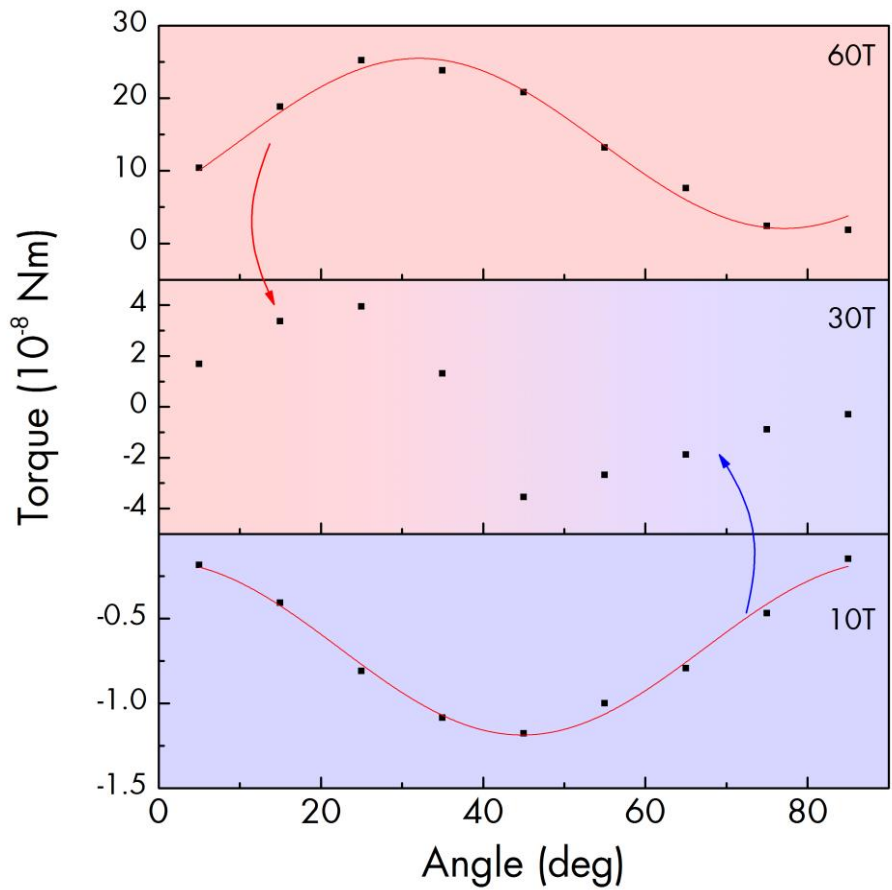

**Fig. 3: Magnetic anisotropy reversal at the quantum limit.** The angle dependence of the torque measured at three different field levels: At 10T in the low field regime, in the high field regime at 60T above the quantum limit for most angles, and at 30T in an intermediate regime. The torque in the low- and high-field regions both follow a $\sin(2\theta)$ dependence, yet the phase shift signals a reversal of the magnetic anisotropy $(\chi_\perp - \chi_\parallel)$ in the two regimes. This change in the magnetic anisotropy is most evident in the intermediate field region, where the system crosses the quantum limit as a function of angle. At low angles, the system follows the high-field anisotropy, then suddenly transitions into the low-field anisotropy at high angles where the quantum limit exceeds 30T, leading to pronounced deviations from the usual $\sin(2\theta)$ behavior. Small deviations from the sign reversal between the 10T and 60T torque may appear due to the eventual cross-over into the low-field regime even at 60T at high angles, and the field dependence of the susceptibility.

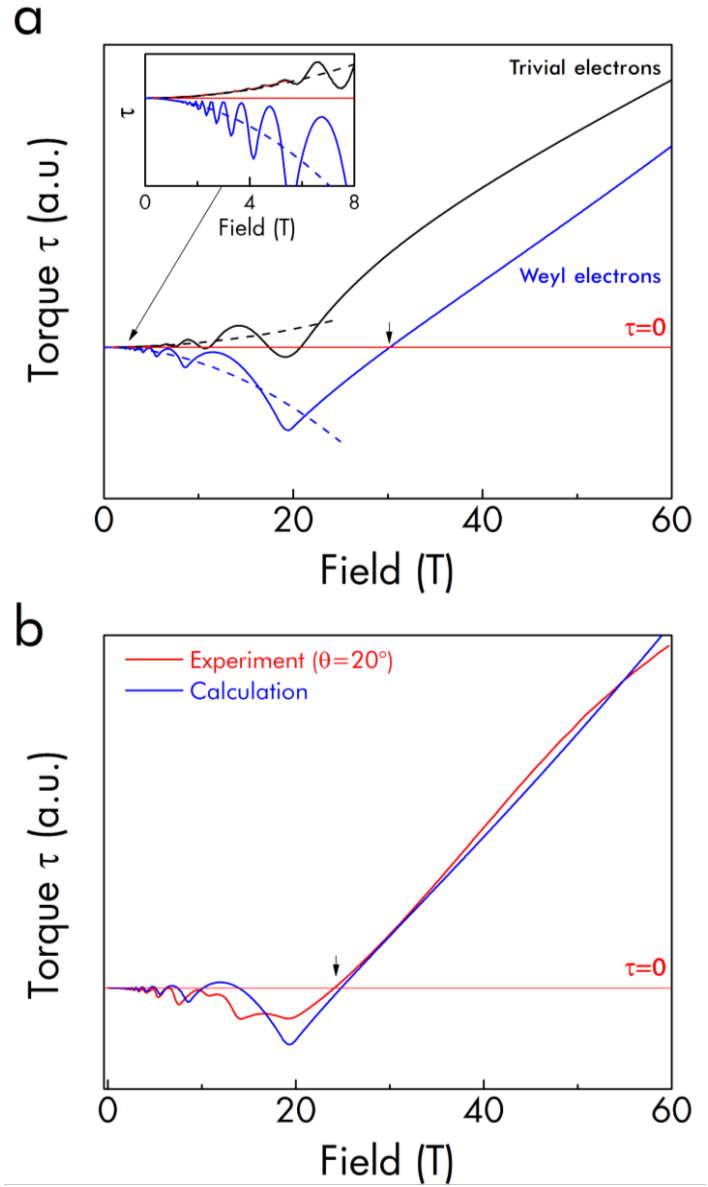

**Fig. 4: Comparison to calculated high field torque.** (a) The torque response form a Weyl metal was calculated using eq. 3 and is contrasted to the expected torque for a trivial metal of similar Fermi surface size. In the trivial metal, the direction of the torque is simply given by the magnetic anisotropy, thus the low-field (dashed line) and the high-field torque point in the same direction. In contrast, the Weyl torque response at low fields is dominated by the charge carriers above the Weyl point. These become magnetically inactive at the quantum limit in the field-independent n=0 state where the weaker but opposite response of the valence states takes over and leads to the sign change of the torque. (b) The calculation matches well with the experimental observation. The only free parameters are the absolute magnitude of the susceptibility, and the momentum cutoff $\Lambda \sim 1 \text{Å}^{-1}$ as described in the main text. Importantly, this calculation is based on an ellipsoidal Fermi surface ignoring the complex band structure of NbAs, which again highlights the topological origin of the torque anomaly.

# References

1.  Vafek, O. & Vishwanath, A. Dirac Fermions in Solids-from High Tc cuprates and Graphene to Topological Insulators and Weyl Semimetals. *Annu. Rev. Condens. Matter Phys.* **5,** 83–112 (2014).

2.  Wehling, T. O., Black-Schaffer, A. M. & Balatsky, A. V. Dirac materials. *Adv. Phys.* **63,** 1–76 (2014).

3.  Hosur, P. & Qi, X. Recent developments in transport phenomena in Weyl semimetals. *Comptes Rendus Phys.* **14,** 857–870 (2013).

4.  Mikitik, G. & Sharlai, Y. Manifestation of Berry's Phase in Metal Physics. *Phys. Rev. Lett.* **82,** 2147–2150 (1999).

5.  Zak, J. Berrys phase for energy bands in solids. *Phys. Rev. Lett.* **62,** 2747–2750 (1989).

6.  Xiao, D., Chang, M.-C. & Niu, Q. Berry phase effects on electronic properties. *Rev. Mod. Phys.* **82,** 1959–2007 (2010).

7.  Geim, A. K. Graphene: Status and Prospects. *Science* **324,** 1530–1534 (2009).

8.  Kane, C. L. An insulator with a twist. *Nat. Phys.* **4,** 348–349 (2008).

9.  Wang, Z., Weng, H., Wu, Q., Dai, X. & Fang, Z. Three-dimensional Dirac semimetal and quantum transport in $Cd_3As_2$. *Phys. Rev. B* **88,** 125427 (2013).

10. Liu, Z. K. *et al.* Topological Dirac Semimetal, $Na_3Bi$. *Science* **343,** 864–867 (2014).

11. Liu, Z. K. *et al.* A stable three-dimensional topological Dirac semimetal $Cd_3As_2$. *Nat. Mater.* **13,** 677–81 (2014).

12. Borisenko, S. *et al.* Experimental realization of a three-dimensional dirac semimetal. *Phys. Rev. Lett.* **113,** 027603 (2014).

13. Huang, S.-M. *et al.* A Weyl Fermion semimetal with surface Fermi arcs in the transition metal monopnictide TaAs class. *Nat. Commun.* **6,** 7373 (2015).

14. Shekhar, C. *et al.* Large and unsaturated negative magnetoresistance induced by the chiral anomaly in the Weyl semimetal TaP. *Arxiv Prepr.* (2015).

15. Shekhar, C. *et al.* Extremely large magnetoresistance and ultrahigh mobility in the topological Weyl semimetal candidate NbP. *Nat. Phys.* 1–6 (2015). doi:DOI:10.1038/NPHYS3372

16. Zyuzin, A. A. & Burkov, A. A. Topological response in Weyl semimetals and the chiral anomaly. *Phys. Rev. B* **86,** 115133 (2012).

17. Shoenberg, D. *Magnetic oscillations in metals*. (Cambridge University Press, 1984).

18. Berry, M. V. Quantal Phase Factors Accompanying Adiabatic Changes. *Proc. R. Soc. A Math. Phys. Eng. Sci.* **392,** 45–57 (1984).

19. Wang, Z. *et al.* Dirac semimetal and topological phase transitions in A 3Bi (A=Na, K, Rb). *Phys. Rev. B* **85,** 195320 (2012).

20. Koshino, M. & Ando, T. Anomalous orbital magnetism in Dirac-electron systems: Role of pseudospin paramagnetism. *Phys. Rev. B* **81,** 195431 (2010).

21. Mikitik, G. P. & Sharlai, Y. V. The phase of the de Haas-van Alphen oscillations, the Berry phase, and band-contact lines in metals. *Low Temp. Phys.* **33,** 439–442 (2007).

22. Mikitik, G. P. & Sharlai, Y. V. Berry phase and de Haas-van Alphen effect in LaRhIn5. *Phys. Rev. Lett.* **93,** 106403 (2004).

23. Goodrich, R. G. *et al.* Magnetization in the ultraquantum limit. *Phys. Rev. Lett.* **89,** 026401 (2002).

24. Ghimire, N. J. *et al.* Magnetotransport of single crystalline NbAs. *J. Phys. Condens. Matter* **27,** 152201 (2015).

25. Weng, H., Fang, C., Fang, Z., Bernevig, B. A. & Dai, X. Weyl Semimetal Phase in Noncentrosymmetric Transition-Metal Monophosphides. *Phys. Rev. X* **5,** 011029 (2015).

26. Luo, Y. *et al.* A novel electron-hole compensation effect in NbAs. *Arxiv Prepr.* (2015). at <http://arxiv.org/abs/1506.01751>

27. Ali, M. N. *et al.* Large, non-saturating magnetoresistance in WTe 2. *Nature* **514,** 205–8 (2014).